\documentclass{aastex}
\usepackage{spr-astr-addons}
\usepackage{url}

\RequirePackage{color}

\begin{document}

\title{Towards a quantum universe}
\slugcomment{Not to appear in Nonlearned J., 45.}
\shorttitle{Towards a quantum universe}
\shortauthors{Jaume Gin\'e}

\author{Jaume Gin\'e\altaffilmark{1}}
\affil{Departament de Matem\`atica, Universitat de Lleida,\\
Av. Jaume II, 69. 25001 Lleida, Spain}


\begin{abstract}
In this short review we study the state of the
art of the great problems in cosmology and their interrelationships.
The reconciliation of these problems passes undoubtedly through the idea of a quantum universe.
\end{abstract}

\keywords{Cosmology, Gravitation theory, Quantum mechanics, General
relativity, large numbers, cosmological constant}


\section{Introduction}

The great challenge of contemporary physics is to reconcile
quantum mechanics, applied at micro cosmos, and general relativity
applied, in general, at macro cosmos. General relativity and
classical electrodynamics equations are invariant under a scale
transformation of time intervals and distances, provided we scale
too the correspondent coupling factors. In particular, the scale
invariance of general relativity was applied to the strong gravity
\cite{SS1,SS2,CPR,SS} that tries to derive the hadron properties
from a scaling down of gravitational theory, treating particle as
black-hole type solutions. Last years, in several works, it was
suggested that also quantum mechanics must be invariant under
discrete scale transformations, see \cite{C}. All suggest that these two
irreconcilable theories, the gravity defined by the General relativity
and the quantum mechanics, can be applied to any scale. Should therefore be
complementary theories that explain the same physical reality.

However, the introduction of the Planck's constant $h$ in the
quantum mechanics defines a very particular scale, at which the
quantum effects must be considered. The quantum equations, as
Schr\"odinger and Dirac ones, are not scale invariants, due to the
presence of $h$. The question that naturally arises is whether it
is really a physical constant at any scale.

The invariance under discrete scale transformations appear from
one of the curious features between particles physics and
cosmology. These features are the possibility of obtaining
cosmological large numbers, as mass $M_U$ radius $R_U$ and age $T$
of the universe, scaling up the typical values of mass $m$, size
$r$ and life time $t$ appearing in particle physics, by the scale
factor $10^{38-40}$. The scale relations are $T/t \thicksim R_U/r
\thicksim (M_U/m)^{1/2} \thicksim \lambda=10^{38-40}$. From here
we can scale $h$ in order to obtain the new constant $\mathcal{H}$
of the new scale invariance of quantum mechanics. From a simple
dimensional analysis we have $\mathcal{H} \thicksim \lambda^3 h$.
The possible meaning of this new constant $\mathcal{H}$ is that
$\mathcal{H}/(2 \pi)$ is the angular momentum of a rotating
universe and this explanation is close to the G\"odel's spin, with
the Kerr limit for the spin, and with the Muradian's Regge-like
relation for galaxies and clusters, see \cite{C} and references
therein. In fact this new constant is $\mathcal{H}\thicksim 10^{120} h$ and
is what is call in \cite{AFa0} the {\it cosmological Planck's constant}. With
this new Planck's constant no large numbers appear at the cosmological level.
In \cite{C} it is also described an intermediate scale
invariance of quantization related to the angular momenta of stars
and close to the Kerr limit for a rotating black hole with mass
around $10^{30} kg$. All these ideas suggest treating the universe
as a single particle, as we shall see later. In fact as a cosmological quantum black hole.
In the following sections we will see that several scaling laws can
explain some of the present cosmological problems.

\section{The Large number coincidence problem}

Hermann \cite{W1,W2} speculated that the observed radius of the
universe might also be the hypothetical radius of a particle whose
energy $m_h c^2$ is equals to the gravitational self-energy of the
electron $Gm_e^2/r_e$, where $m_h$ is the mass of the hypothetical
particle, $m_e$ and $r_e$ the mass and the radius of the electron.
This was the beginning of the large number coincidence problem.
Hence, we have
\[
\frac{R_U}{r_e} \thicksim
 \frac{r_h}{r_e}=\frac{m_e}{m_h}=\frac{m_e}{\frac{Gm_e^2}{c^2r_e}}=\frac{c^2
r_e}{G m_e}=\frac{e^2}{4 \pi \varepsilon_0 G m_e^2} \approx
10^{42},
\]
where we have that $r_e=e^2/(4 \pi \varepsilon_0
m_e c^2)$ and $r_h=e^2/(4 \pi \varepsilon_0 m_h c^2)$. This
coincidence was further developed by \cite{E} who
related the above ratios with $N$, the estimated number of charged
particles in the universe.
\[
\frac{e^2}{4 \pi \varepsilon_0 G m_e^2} \thicksim \sqrt{N} \approx
10^{42}.
\]
Eddington obtained the most intriguing relation between the
present number of baryons in the universe, known as the Eddington
number, and the squared ratio of the electric to the gravitational
force between the proton and the electron.
\[
\frac{F_e}{F_g}=\frac{e^2}{4 \pi \varepsilon_0 G m_e m_p}
\thicksim \frac{cT}{r_e} \approx 10^{40},
\]
where $m_p$ is the proton mass and $T$ is the age of the universe.
This coincidence between large numbers can also be expressed in
the alternative form
\begin{equation}\label{EW}
\hbar^2 H_0 \thicksim G m_n^3 c \, ,
\end{equation}
where $m_n$ is the nucleon mass, $H \equiv \dot{a}/a$ is the
Hubble parameter and $H_0$ is its present value and $a(t)$ is the
scale factor, see \cite{MC}. This approximate identity is called the {\it
Eddington-Weinberg relation}. The Hubble parameter is not a
constant and varies as the inverse of the cosmological time $t$ in
the standard Friedmann-Robertson-Walker (FRW) cosmology. This fact
led \cite{Dir,Dir2} to speculate the hypothesis that
Newton's constant $G$ must depend on time as $H$, i.e. $G
\thicksim 1/t$, so that relation (\ref{EW}) remains always valid.
This fact is incompatible with the experimental bounds that exist
on time variation of $G$, see \cite{DGT,MC,WND}. Hence, the coincidence
(\ref{EW}) is only valid in this epoch. In \cite{F} it was
resolved the large number coincidence problem using scaling laws
from the standard cosmological model. We reproduce here the
arguments. From the scale relations (that do not constitute a
coincidence problem)
\[
\left ( \frac{M_U}{m_n} \right)^{1/2} \thicksim \left (
\frac{m_P}{m_n} \right)^2 \thicksim \left ( \frac{M_U}{m_P}
\right)^{2/3},
\]
where $m_P$ is the Planck mass, we obtain $M_U \thicksim
m_P^4/m_n^3$ where $M_U= \Omega_m (4 \pi/3) R_U^3 \rho_c = (4
\pi/3) R_U^3 \rho_m$ is the observable mass of the universe. The
Hubble parameter in a universe with zero curvature is related with
the average total energy density $\varepsilon$ by
\begin{equation}\label{es}
H^2= \frac{8 \pi G \varepsilon}{3 c^2},
\end{equation}
and during the matter-dominance the total energy is
$\varepsilon=c^2 \rho_m$. Therefore the mass of the universe is
equals
\begin{equation}\label{eq}
M_U= \frac{4 \pi}{3}R_U^3 \rho_m =  \frac{R_U^3 H^2}{2 \, G}.
\end{equation}
Taking into account that $H \thicksim c/R_U \thicksim 1/T$
equation (\ref{eq}) gives the scaling law
\begin{equation}\label{exp}
GM_U \thicksim c^2 R_U.
\end{equation}
Expression \eqref{exp} was obtained by \cite{W}, \cite{WR}, \cite{Sc},
\cite{BD,Di} and also by \cite{As,As2} in different contexts. Another form to obtain
equation (\ref{exp}) is applying the classical Mach's principle by
requiring that the self-energy of a body is given by the
gravitational energy of interaction of a body with the whole
universe:
\[
mc^2= \frac{GmM_U}{R_U}.
\]
Substituting this scaling law \eqref{exp} in the expression
$M_U=m_P^4/m_n^3$ and remembering that the Planck mass is $m_P=
\sqrt{\hbar c/G}$ we have
\[
\frac{c^2 R_U}{G}= \frac{m_P^4}{m_n^3}=\frac{\hbar^2 c^2}{G^2
m_n^3},
\]
and from here the Eddington-Weinberg relation (\ref{EW}).

\section{The Cosmic coincidence problem}

In an expanding universe with scale factor $a(t)$, where $t$ is
the cosmological time, $\Lambda$ is a constant while the matter
density $\rho_m$ decreases with $a^3$. However, the observed
energy density of matter $c^2 \Omega_m \rho_c$ is so close to the
vacuum energy density attributed to the cosmological constant
$\Lambda$, given by $\varepsilon_{vac}= 3 \Lambda c^2 / (8 \pi
G)$. This coincidence is known as the cosmic coincidence problem
and may be expressed as
\[
\rho_m = \Omega_m \rho_c \thicksim \frac{3 \Lambda}{8 \pi G}.
\]
As in the case of the large number coincidence, this coincidence
occur only in this epoch. We are going to see that the cosmic
coincidence problem is a consequence of the large number
coincidence and due to the fact that we are in the era of vacuum-dominance. If
we assume that the present evolution of the universe is dominated
by the cosmological constant $\Lambda$, as corroborated by
observation \cite{TZH}, we can set $H_0 \thicksim \Lambda^{1/2}$.
The continuous transition from the matter-dominance given by
equation (\ref{es}) to our era of vacuum-dominance gives the
cosmic coincidence
\[
H_0^2 \thicksim \frac{8 \pi G \rho_m}{3} \thicksim \Lambda.
\]

\section{The Cosmological constant problem}

If $\Lambda$ originates from the vacuum quantum fluctuations, its
theoretically expected value has order of $l_p^{-2}$ where $l_p
\equiv \sqrt{\hbar G/c^3} \approx 10^{-35} m$ is the Planck
length, see \cite{Wei}. That is, 122 orders of magnitude greater
than the observed value $\Lambda \approx 10^{-52} m^{-2}$, see
\cite{TZH}. This huge discrepancy is known as the cosmological
constant problem and it is an open problem nowadays, see for
instance \cite{Wei,HS}.

However we can get for the cosmological constant $\Lambda$ one
scaling law that also explains the cosmic coincidence, see
\cite{F,F2}. Putting the condition $H_0 \thicksim \Lambda^{1/2}$ in
the large number coincidence (\ref{EW}) we have
\begin{equation}\label{Zel}
\Lambda \thicksim \frac{G^2 m_n^6 c^2}{\hbar^4}.
\end{equation}
Equation (\ref{Zel}) is essentially the same scaling law derived
by \cite{Z}, from considerations of field theory and
empirical arguments. This form to derive equation (\ref{Zel}) was
first made by \cite{Ma}, who takes relation (\ref{EW}), as
well as the present dominance of the cosmological constant over
the density of matter. Taking into account that the Compton
wavelength of the nucleon is $\lambda_n=h/(m_n c)$ and the scale
relation
\begin{equation}\label{sca}
\left ( \frac{M_U}{m_n} \right)^{1/2} \thicksim
\frac{R_U}{\lambda_n},
\end{equation}
from (\ref{Zel}) we obtain the scaling laws
\begin{equation}\label{Z}
c^2 \Lambda \thicksim \frac{G^2 m_n^2}{\lambda_n^4}  \thicksim
\frac{G^2 M_U^2}{R_U^4}.
\end{equation}
This scaling law says that the energy density associated to the
cosmological constant may be scaled to the gravitational energy of
the nucleon mass confined to a sphere whose radius is the Compton
wavelength of the nucleon and to the gravitational energy of the
universe of mass $M_U$ and whose radius is $R_U$. This is the generalization
of the \cite{Z} equation (\ref{Zel})
to the cosmological level
\begin{equation}\label{Zel2}
\Lambda \thicksim \frac{G^2 M_U^6 c^2}{\mathcal{H}^4},
\end{equation}
with the introduction of the cosmological Planck's constant $\mathcal{H}$
satisfying $R_U=\mathcal{H}/ (M_U c)$ and the generalization of the Eddington-Weinberg relation (\ref{EW})
\begin{equation}\label{EW2}
\mathcal{H} H_0 \thicksim G M_U^3 c \, ,
\end{equation}
assuming that the present evolution of the universe is dominated
by the cosmological constant $\Lambda$ and then we have $H_0 \thicksim \Lambda^{1/2}$.
These generalizations are also obtained in \cite{AFa0,AFa}. Moreover the cosmological constant problem
is solved with the introduction of the cosmological Planck's constant $\mathcal{H}$
because now $\Lambda_c$ originates from the cosmological vacuum quantum fluctuations, has the value of order $L_p^{-2}$ where $L_p
\equiv \sqrt{\mathcal{H} G/c^3} \approx 10^{26} m$ is the cosmological Planck
length, and we obtain $\Lambda_c \approx 10^{-52} m^{-2}$ which agrees with
the observed value. In fact, this cosmological Planck length $L_p$ is of order of the radius of the universe $R_U$. Hence we have $R_U^2 \thicksim \mathcal{H} G/c^3$, that taking into account $R_U=\mathcal{H}/ (M_U c)$ we
reobtain the equality (\ref{exp}) that relates $M_U$ with $R_U$.
In resume we have the following identities that define the cosmological scale $R_U=GM_U/c^2$, the cosmological Compton wavelength $\bar{\lambda}_c = \mathcal{H}/(M_U c)$, the new cosmological constant $\Lambda_c \thicksim L_p^{-2} \thicksim R_U^{-2}$ and it is satisfied that $\Lambda_c \mathcal{H}= c^3/G$. Hence we have two important scales, the micro scale called Planck scale and the macro scale given by the cosmological scale that suggest the scale relativity invariance introduced by \cite{N0}.

\section{The critical acceleration coincidence}

The observed motions of clusters of galaxies and material within
galaxies may be interpreted to indicate that the laws of dynamics
deviate from Newtonian models at accelerations smaller than some
critical acceleration $a_0 \approx 10^{-10} m s^{-2}$, see
\cite{Mi1}. The Hubble acceleration $c H_0$ is of the same order
only in this epoch. This coincidence $a_0 \thicksim c H_0$ is well
known from the first works of \cite{Mi1} see also \cite{F}. This coincidence
is justified in \cite{G} and \cite{G3} by different arguments.
Substituting $H_0 \thicksim \Lambda^{1/2}$ the coincidence takes
the form $a_0 \thicksim c \Lambda^{1/2}$ and taking into account
the scaling law (\ref{Z}) we obtain
\[
a_0 \thicksim \frac{Gm_n}{\lambda_n^2}.
\]
Hence, the critical acceleration is scaled to the characteristic
gravitational acceleration of the nucleon mass at its Compton
length. Moreover, taking into account the scale relation
(\ref{sca}) we have that
\begin{equation}\label{sts}
a_0 \thicksim \frac{G M_U}{R_U^2}.
\end{equation}
Hence, the critical acceleration is scaled to the characteristic
gravitational acceleration of any body in our universe due to the
all the rest of the mass of the universe. This interpretation of
the critical acceleration appears in \cite{G} in the context of a
implementation of the inertia Mach's theory. In \cite{T1,T2} it is found that 
identity (\ref{sts}) is invariant at any scale because is satisfied by
the hadrons, the electrons, the nucleus, the globular clusters, the galaxies,
the clusters of galaxies, the universe as a whole and others physical situations.

\section{The cosmic acceleration problem}

The standard candle observations of type Ia supernovae give a
cosmic acceleration with a positive rate, which implies the
introduction of the cosmological constant in the cosmological
models. Hence, the expansion of the universe is accelerating, see
\cite{Ri}. This acceleration states the cosmic acceleration problem.
The question is what causes this acceleration?

It is clear that the introduction of the cosmological constant
give as a consequence that the universe is accelerating. However,
what is the nature of this cosmological constant introduced?

There are essentially two ways of introducing the cosmological
constant or the dark energy. The first one is changing gravitation
with $f(R)$ gravity models, Scalar-tensor models, braneworld
models, etc. The second one is changing matter with the
quintessence, K-essence, tachyons, Chaplygin gas, phantom field,
etc.

We have seen that the Eddington-Weinberg relation (\ref{EW}) is
only valid in this epoch. However, the strong version of the
cosmological holography principle also implies equation
(\ref{EW}), see \cite{C2}, without any additional assumption as
the dominance of the cosmological constant $\Lambda$. Hence, in
this case the relation (\ref{EW}) is valid for any cosmological
time. The derivative respect to the time of the relation
(\ref{EW}) gives $\dot{H}=0$, because the variation of $G$ is
incompatible with the observations. Now, we recall the definition
of the deceleration parameter $q= - a \ddot{a}/ \dot{a}^2$ and its
relation with the Hubble parameter
\[
\dot{H}=-(1+q) H^2.
\]
Therefore, the strong version of the cosmological holography
principle implies that $q \approx -1$ in order to obtain
$\dot{H}=0$. This value of the deceleration parameter is also
found in the context of the modified Newtonian theory (MOND) in
\cite{G4}, when we evaluate the recessional acceleration $a_r(t)= -q H v_r$
for the objects receding from us at a rate faster than the speed
of light and compare with the value of the constant acceleration $a_0=H_0c$.
In this case the Hubble law is applied for close distances assuming the same
behavior at first order for largest observable distance. 

\section{The quantum universe}

We have seen the existence of several scaling laws that explain
some of the present cosmological problems. However, the origin of the dark energy and dark matter are still open problems.

In \cite{AFa} it is given a necessary and sufficient condition for an object of any mass $m$ to be a quantum black hole generalizing the results obtained for the cosmological scale.  This generalization is established by the following identities that define a quantum black hole for each $m$ and a new scale. The first is $r_m=G m /c^2$, where $r_m$ is the gravitational radius, the generalized Compton wavelength $\bar{\lambda}_m = h_m/(m c) \thicksim r_m$, 
where $h_m$ is the generalized Planck's constant, the $\Lambda_m \thicksim r_m^{-2}$ and it is satisfied that $\Lambda_m h_m= c^3/G$. This generalization is also justified by the described intermediate scale
invariance of quantization for a rotating black hole with certain mass, see \cite{C}.
Hence in \cite{AFa,FA} is adopted the idea that the universe is a quantum black hole and therefore it is possible to define, following the \cite{Ha} formulation, the entropy of the universe as a quantum black hole
\[
S= \frac{4 \pi k_B}{\hbar c} G M^2 = \pi k_B \left( \frac{R_U}{l_p}\right)^2 \approx 10^{122} k_B,
\]
which is in accordance with the current value found by \cite{EL}. In \cite{FA} it is computed the conjugate black hole of the universe that is identified with
the quantum of the gravitational potential field and the bit. Besides, the information-entropy relation, based on the bit, the \cite{P1,P2} proposal that gravity has an entropic or thermodynamic origin, and the \cite{V} interpretation
of gravity as an emerging entropic force, gives a hope to unify gravity with quantum theory.

The idea of Alfonso-Faus \& Fullana reinforces the relationships between the constants of atomic physics and the constants of the Universe, as we have described in the text, see also \cite{H0,Din}. In \cite{H0} three interesting relations are presented. The first one connects the Compton wavelength of a pion and the dark energy density of the universe; the second one connects the Compton wavelength of a pion and the mass distribution of non-baryonic dark matter in a galaxy; the third one relates the mass of a pion to fundamental physical constants and cosmological parameters which has as particular case the Eddington-Weinberg relation (\ref{EW}) but for the pion mass. The importance of the pions (instead of the nucleon mass) is due to ``virtual'' pions, which are, accoding to quantum field theory, an inherent part of vacuum fluctuations and as a simple particles (quark pairs) dominate the quantum vacuum.
We recall that pions are the subatomic particles that describe the interaction between nucleons. Under this scenario, each nucleon is continuously emitting and reabsorbing virtual pions, which surround it like a swarm. Moreover correct value of mass to put in the identity (\ref{Zel}) according to the observed value of $\Lambda$ is about 1/20 times the proton mass or about 80 times the electron mass and is about one third the pion mass, see \cite{S1}. Therefore the pions must dominate the quantum vacuum fluctuations that contribute to the value of the cosmological constant. In \cite{Din} it is derived the values of the baryon density parameter, the Hubble constant, the cosmic microwave background temperature and the helium mass fraction in excellent agreement with the the most recent observational data.

Following the idea of a quantum vacuum fluctuations, with virtual particles flashing in and out of existence, in \cite{S1,S2} it is showed that the vacuum fluctuations effectively supplies a vacuum energy pressure which is of the right order of magnitude to explain dark energy.
The key idea of the Santos works is the two-point correlation function of vacuum fluctuations gives the correct contribution of Dark energy, and this relies upon the disappearance of the correlation within the Planck length which solves the cosmological constant problem.

The following exciting papers papers can shed light on the nature of the dark matter and the solution of the dark matter problem.  In \cite{V1} showed that, from CPT invariance of the general relativity, the sign of the gravitational force between matter and antimatter is reversed (anti-gravity). This is a controversial result which is being analyzed and discussed, see for instance \cite{Cab,Cro} and \cite{V2}.

Based in the anti-gravity (that a particle and its antiparticle have the gravitational charge of the opposite sign) \cite{H2,H3} consider that the quantum vacuum may be considered as a fluid of virtual gravitational dipoles. In such a way that when we place a gravitational mass in a quantum vacuum will induce a polarization of the quantum vacuum, in the same way that a charge induces polarization in a surrounding dielectric medium. In the case of gravitation, we would expect to find more virtual particles close to a gravitating object, and more anti-particles at much greater distance. This would mean that, in a galaxy for example, the apparent gravitational attraction of the body is an increasing function of distance out to some critical value.
Following this hypothesis, Hajdukovic present the first indications that dark matter may not exist and that the phenomena for which it was invoked might be explained by the gravitational polarization of the quantum vacuum by the known baryonic matter. The best developed alternative to particle dark matter is the Modified Newtonian Dynamics (MOND) \cite{Mi1}, but we witness a violation of the fundamental law of gravity and has still fundamental problems with the observational data, see for instance \cite{G4,G5} and references therein. However in the Hajdukovic model the distribution of vacuum polarization will depend on the distribution of matter, so the apparent extra acceleration towards the center of mass will vary from one object to another, and as a function of position within the object, see \cite{H1}. Moreover the consequences of the model can be tested, see \cite{H3}, where some phenomena partially explained by dark matter and theories of modified gravity are understood in the framework of the gravitational polarization. Moreover the theory presented in \cite{H2,H3} is not a support to MOND although there is a critical gravitational filed which corresponds to the maximal gravitational polarization density.\\
  
The final conclusion is that is needed a quantum gravitational theory with a quantum granulation of space-time and in this new framework the presented papers have given us grounds to hope that both dark energy and dark matter will find their natural explanation as simply naturally-arising quantum vacuum phenomena.

\acknowledgments
The author is partially supported by a MICINN/FEDER grant number
MTM2011 -22877 and by a Generalitat de Catalunya grant number 2009SGR 381



\end{document}